# Nanoparticle lattices with electric and magnetic resonances [Prospective]

Viktoriia E. Babicheva[1,*] and Andrey B. Evlyukhin[2]

[1]College of Optical Sciences, University of Arizona, 1630 E. University Blvd., Tucson, AZ 85721

[2]ITMO University, 49 Kronverksky Ave., St. Petersburg, 197101, Russia

*email: vbab.dtu@gmail.com

**Abstract**

Lattice resonances in nanoparticle arrays recently have gained a lot of attention because of the possibility to produce spectrally narrow resonant features in transmission and reflection as well as significantly increase absorption in the structures. Most of the efforts so far have been put to study these lattice resonances in dipole approximation. However, the recent research shows that higher multipoles not only produce resonant feature but are also involved in cross-coupling, affect each other, and induce a magnetoelectric response. In this *Prospective*, we review the recent achievements in studying of interplay and coupling of different multipoles in periodic nanoparticle arrays and share our vision on further progress of the field.

## I. Introduction

Nanoparticles exhibit a variety of interesting optical properties [1]. Plasmonic particles support resonances of localized surface plasmons, which result in high field concentration in the proximity of the particles and more efficient manipulation of light at the nanoscale. Nanoparticle assembles, like oligomers and clusters, support a broad range of resonances [2], their interplay causes sharp features in the spectra, including so-called Fano resonances [3], and consequently can be utilized in functional optical elements and metasurfaces. It has been shown that subwavelength plasmonic structures can enhance light-matter interaction [4,5] and open up possibilities of a wide range of applications such as optical antennas [1], photovoltaics [6,7], scattering-type near-field optical microscopy [8], and others.

Particles arranged in periodic lattices enable even more fascinating properties, and the most prominent effects happen when the period of the array is comparable with the wavelength of nanoparticle resonance. Being in the proximity of single-particle resonance maximum, these lattice resonances strongly modify the spectral profile [9-15], but for an offset to the red part of the single-particle resonance, the lattice resonances appear as additional separate features. Field enhancement and more efficient scattering that results from lattice resonance excitations [16-21] can find applications in sensors [22], nanolasers [23], light harvesting devices [24,25], modulators [26], and others. The broad variety of enabled functionalities are highlighted in the recent review [27], dipole coupling of multiple particles in the cell [28], and multipolar interactions in the surface-lattice resonances in two-dimensional arrays of spheres [29,30]. Different nanoparticles have been demonstrated to enhance lattice resonances in orthogonal and parallel coupling between particles [31-35]. In this Prospective, we are focused on the overview of effects behind the pronounced lattice features, lattices of uncoupled multipoles considered so far (e.g. [36, 37]), and recently demonstrated multipole coupling in the infinite arrays even under the normal incidence of external light waves [38]. We envision how the processes of electric and magnetic multipoles interplay, cross-coupling, and resonance induction will enrich the field by bringing both fundamental understandings of the effects and opportunities for practical applications.

## II. Electric-dipole lattice resonances

Dipole coupling in one- and two-dimensional nanoparticle arrays can produce collective lattice resonances, and their wavelengths are determined by the lattice periods. In such electric dipole (ED) approximation, one need taking into account only dipole moments of the nanoparticles oriented perpendicular to the lattice direction where the period is comparable to the effective wavelength of the nanoparticle resonance. Under normal incidence of light, every identical spherical nanoparticle arranged in the infinite periodic array have the same effective electric dipole moment $p_0$ calculating from the equation:

$$p_0 = \alpha_p E_x(\mathbf{r}_0) + \frac{\alpha_p k_0^2}{\varepsilon_0} S_{pp} p_0, \qquad (1)$$

where $\alpha_p$ is the ED polarizability, $S_{pp}$ is the ED sum accounting for the electromagnetic interaction between the nanoparticle array, $E_x(\mathbf{r}_0)$ is the electric field of an incident light wave polarized along *x*-axis located on the array plane, $k_0$ is the wave number in a vacuum, and $\varepsilon_0$ is the vacuum permittivity. For more details, we refer to Refs. [14, 36, 37], and in the Prospective, we consider the only normal incidence of light. Multipole particle polarizabilities, for instance, $\alpha_p$ for ED, can be calculated from Mie theory coefficients and the approach is shown for dipoles and quadrupoles in Refs. [36, 37].

One can show that the effective polarizability of the particle defined as $\alpha_p^{\text{eff}} = p_0 / E_x$ exhibit singularity at the wavelength close to the period of the structure. In particular, for the case of $E_x$ along the *x*-axis and the wavelength close to Rayleigh anomaly $\lambda \approx \lambda_{\text{RA,eff-1}} = D_y$ (transverse period, denoted below as $D_t$), the lattice resonances significantly modify the resonance profile of the particles in the array in comparison to a single particle. At the same time, at the wavelength close to another Rayleigh anomaly $\lambda \approx \lambda_{\text{RA,eff-2}} = D_x$ (transverse period, denoted below as $D_p$), only slight changes of the resonance profile take place.

The case of lattice resonances in dipole approximation has been the most extensively studied so far [9-15]. Being mainly observed in the array of plasmonic nanoparticles, it is often referred as "plasmonic" or "surface" lattice resonances. However, as we show below, the resonant feature can be found in many other optical structures and does not necessarily require plasmonic particles, and the deeper study brings us to a much richer variety of effects.

### III. Magnetic-dipole lattice resonances

For the particles supporting magnetic dipole (MD) resonance, like silicon nanospheres [36,41,42] and other simple shapes [43,44] or core-shell nanoparticles, one can show the possibility of corresponding lattice effect, and the effective magnetic moment $m_0$ of identical particles in the arrays can be found as:

$$m_0 = \alpha_m H_y(\mathbf{r}_0) + \alpha_m S_{mm} m_0, \qquad (2)$$

where $\alpha_m$ is the MD polarizability, $S_{mm}$ is the MD sum accounting for the electromagnetic interaction between the nanoparticle array, and $H_y(\mathbf{r}_0)$ is the magnetic field of normally incident light wave linearly polarized (with respect to the electric field) along the x-axis. For more details see e.g. Ref. [36]. Thus, similar to EDs, lattice resonances of magnetic counterparts can be spectrally varied, and their wavelength is close to the Rayleigh anomaly $\lambda \approx \lambda_{\text{RA,eff-2}} = D_p$ (Fig. 1). As has been shown in the initial work [36], the lattice of particles with ED and MD responses can be described by the system of Eqs. (1) and (2), and the electric (magnetic) dipoles can be considered independently from magnetic (electric) counterparts, that is uncoupled.

We recently studied two-dimensional periodic arrays of silicon and core-shell nanoparticles that support lattice resonances due to the ED and MD resonances of the nanoparticles [40]. We showed a possibility of achieving a full overlap between the ED and MD nanoparticle resonances adjusting lattice periods independently in each mutual-perpendicular direction. In this way, one can realize the resonant lattice Kerker effect, that is resonant suppression of the backward-scattered waves (reflectance) from the array. The strong suppression of light reflectance of the structure is appeared due to destructive interference between light scattered by EDs and MDs of every nanoparticle in the backward direction with respect to the incident light wave. The resonant lattice Kerker effect based on the overlap of both ED and MD lattice resonances as well as an experimental proof of independent resonance control [45] have also been demonstrated.

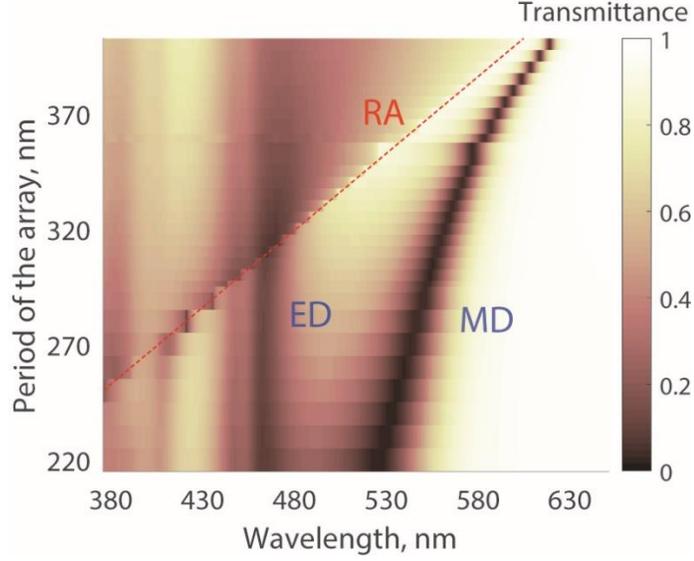

Fig. 1. The array transmittance and the change of MD peak resonance wavelength for different periods $D_p$. Silicon nanoparticles have R = 65 nm, and the arrays are in a dielectric matrix with refractive index n = 1.5. Red line shows the wavelength of Rayleigh anomaly (RA). The transverse period is fixed to $D_t$ = 220 nm.

**IV. Electric-quadrupole lattice resonances**

The particles of larger size and/or complex shape support higher multipoles, and lattice resonances are not limited by the dipole approximation. The case of particles with ED and EQ resonances has been considered in Ref. [37]. Similar to the dipole array, the quadrupoles can be described by the equation:

$$Q_0 = \frac{\alpha_Q i k_S E_x(\mathbf{r}_0)}{2} + \frac{\alpha_Q}{2\varepsilon_0} S_{QQ} Q_0, \qquad (3)$$

where $\alpha_Q$ is the electric quadrupole polarizability, $k_S$ is the wave number in the surrounding medium, $S_{QQ}$ is the electric quadrupole sum accounting for the electromagnetic interaction between the nanoparticle array, and $Q_0$ is the matrix element in the particle electric quadrupole moment $\hat{Q} = Q_0(\hat{x}\hat{z} + \hat{z}\hat{x})$.

The work [37] has outlined an idea that lattice resonances can be achieved with higher multipole resonances, which provide broader opportunities for control of resonant features in the structures and designing optical elements based on them. Alike to the case of EDs and MDs, it has been shown that the lattice of EDs and EQs can be described by Eqs. (1) and (3), and they are not coupled to their counterparts.

**V. Multipole coupling in the lattices**

The situation drastically changes in the case when the lattice includes a couple of non-zero MD and EQ moments. The equation system describing the lattice with dipole and quadrupole moments under normal incidence of the external light waves is the following:

$$m_0 = \alpha_m H_y(\mathbf{r}_0) + \alpha_m \left[ S_{mm} m_0 + \frac{k_0 c}{i} S_{mQ} Q_0 \right], \qquad (4a)$$

$$Q_0 = \frac{\alpha_Q i k_S E_x(\mathbf{r}_0)}{2} + \frac{\alpha_Q}{2\varepsilon_0} \left[ \frac{i k_0}{c} S_{Qm} m_0 + S_{QQ} Q_0 \right], \qquad (4b)$$

where $S_{mQ}$ and $S_{Qm}$ are the sums of cross-effects of EQ on MD and MD on EQ, respectively [38].

As has been shown recently, the terms $S_{mQ}$ and $S_{Qm}$ are not equal to zero indicating a cross-coupling of MD and EQ in the lattices [38]. An example of cross-coupling between multipoles is shown in Fig. 2a,b. Because of the symmetry of equations with respect to electric and magnetic fields, one can predict the similar effect for ED and MQ: $S_{pM}$ and $S_{Mp}$ are expected to be non-zero and the multipoles in lattice induce each other's resonances in the spectral proximity to Rayleigh anomaly.

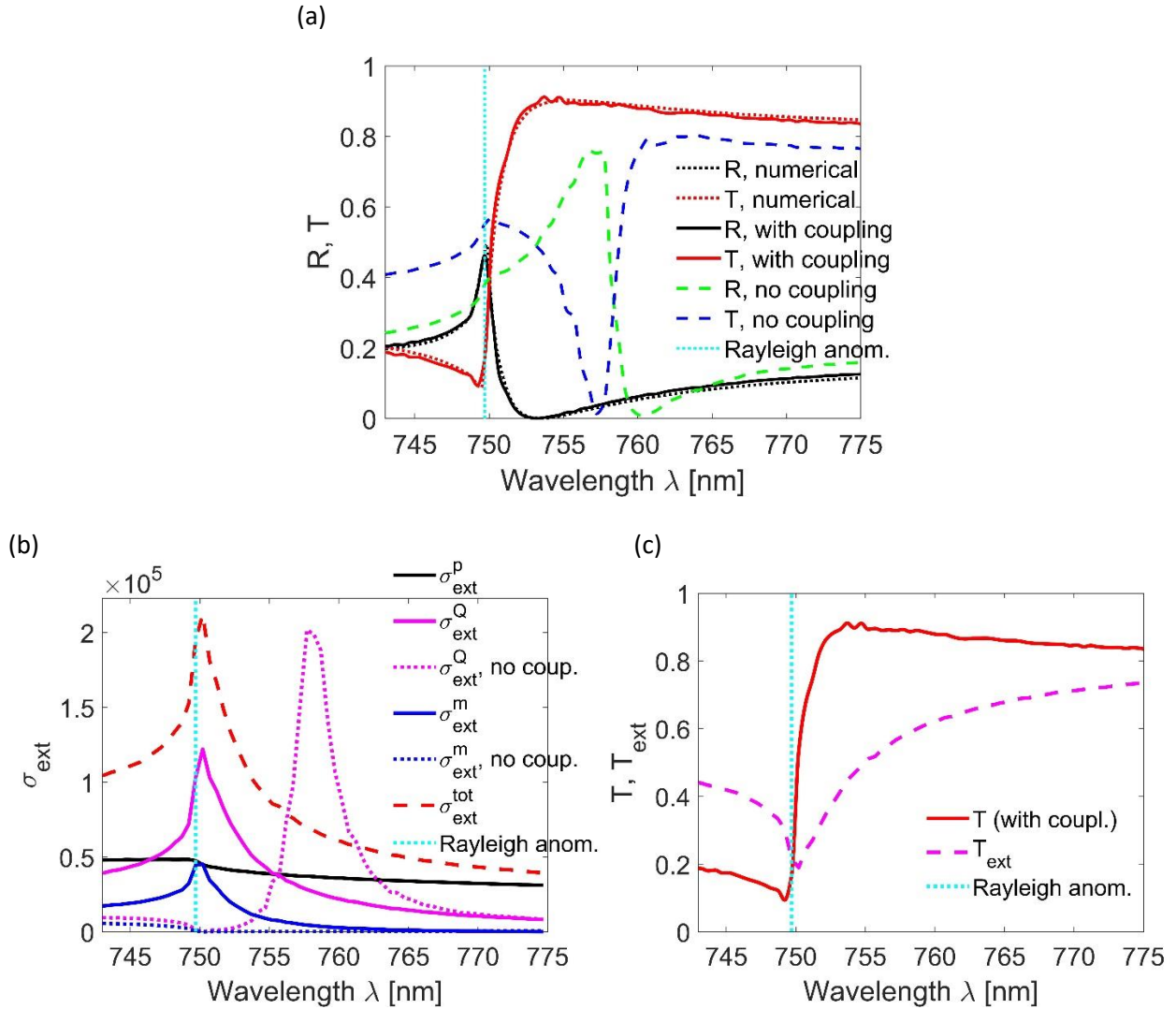

Fig. 2. (a) Comparison of numerical and semi-analytical calculations of reflectance $R_0$ and transmittance $T_0$ for the cases without EQ and MD coupling ("no coupling" in the legend) and with EQ and MD coupling ("with coupling" in the legend). Agreement between semi-analytical calculations and numerical simulations is striking good, and one can see that coupling between EQ and MD multipoles must be taken into account for accurate calculations of resonant profiles. (b) Extinction cross-sections: EQ and MD resonance are excited at the wavelength of the Rayleigh anomaly, and these moments make a detrimental contribution to the total extinction cross-section. (c) Transmittance $T_0$ and $T_{\text{ext}}$. The nanoparticle array with gold spheres of radius $R$ = 100 nm, gold permittivity is taken from experiment [39], and the array is in a dielectric matrix with refractive index n = 1.47. Periods are $D_p$ = 510 nm and $D_t$ = 250 nm.

The array transmission $T_0$ (calculated for the zeroth diffraction order and accounting for all three moments and their interference) is defined as [38]

$$T_0 = \left|1 + \frac{ik_S}{2S_L}\left[\frac{1}{\varepsilon_0\varepsilon_S}\alpha_p^{\text{eff}} + \alpha_m^{\text{eff/coup}} + \frac{k_0^2}{12\varepsilon_0}\alpha_Q^{\text{eff/coup}}\right]\right|^2, \qquad (5)$$

where $\alpha_p^{\text{eff}} = p_0/E_x$, $\alpha_m^{\text{eff/coup}} = m_0/H_y$, and $\alpha_Q^{\text{eff/coup}} = 2Q_0/(ik_S E_x)$ are effective polarizabilities of ED, MD, and EQ respectively, defined from the Eqs. (1) and (4) taking into account coupling between MD and EQ moments; $S_L = D_p D_t$ is the area of the unit cell; $\varepsilon_S$ is the surrounding medium permittivity.

Signal extinction in the array $T_{\text{ext}}$ is defined as

$$T_{\text{ext}} = \exp\left[-\left(\sigma_{\text{ext}}^p + \sigma_{\text{ext}}^m + \sigma_{\text{ext}}^Q\right)/S_L\right], \qquad (6)$$

where $\sigma_{\text{ext}}^p = (k_0/\sqrt{\varepsilon_S})\text{Im}[\alpha_p^{\text{eff}}]$, $\sigma_{\text{ext}}^Q = (k_0^2 k_S/12)\text{Im}[\alpha_Q^{\text{eff/coup}}]$, and $\sigma_{\text{ext}}^m = (k_0/\sqrt{\varepsilon_S})\text{Im}[\alpha_m^{\text{eff/coup}}]$ are extinction cross-sections of ED, MD, and EQ respectively. We note that there is no direct explicit relation between the array transmission $T_0$ and extinction of signal in the array $T_{\text{ext}}$, but both $T_0$ and $T_{\text{ext}}$ are strongly affected by the lattice effect and have pronounced feature at the wavelength of Rayleigh anomaly (Fig. 2c).

**IV. Outlook**

In this work, we have shown a dipole-quadrupole model for infinite arrays with identical nanoparticles under normal incidence of light and discussed a coupling between electric quadrupole and magnetic dipole moments resulting in a resonant feature in the proximity to Rayleigh anomaly. Typically, the lattice resonances are narrow in comparison to dipole resonances of a single particle, and because of the high sensitivity of collective resonances to the optical properties fo surrounding environment, lattice resonances can be used in sensing applications. For the realistic nanoparticle array with a finite number of particles, parameter deviations, experimental uncertainties, imperfections, and defects, one can expect that lattice resonances will decrease and broaden, or possibly smear out. Previous works on plasmonic nanoparticle arrays had shown the lattice resonances appear already in the array of about 50 particles and variations in light incidence angle close to the realistic experiments [37,46]. Furthermore, lattice resonances and lasing enabled by them have also been shown to persist even upon removing 99% particles from the array [47]. We would like to emphasize that the considered effect of lattice resonance excitations is expected to be possible for experimental observation as one can predict from the earlier experimental study of lattice resonances in transverse polarizations [48] which has been a motivation of our earlier study [38].

The recently demonstrated effect of coupled-multipole lattice resonances has the following implications:

1. In the case of small multipole excitations of single particles, their arrangement in the periodic array may significantly enhance the multipole's response, induce a magnetoelectric coupling, and result in the resonant spectral features. For instance, the gold nanospheres with the radius down to 80 nm appear to have only ED resonance, but the particle arrangement in the lattice results in EQ collective resonance [38].

2. In the case when only one multipole in the couple ED-MQ or MD-EQ is pronounced and another one is insignificant yet non-zero, the small multipole excitation can be enhanced by the counterpart. For instance, it has been shown that in the array of gold nanospheres with a radius of 100 nm, the lattice resonance is affected not only by EQ but also lattice-induced MD (see Fig. 2 and Ref. [38], compare calculations with and without taking into account multipoles coupling to the *ab initio* full-wave simulations). Thus, lattice resonances induce a magnetic response from the array of particles without a pronounced magnetic moment.

3. Lattice-induced multipole resonances can be strong enough to have their contribution to the reflection and transmission of the array comparable to the one of single-particle dipole resonance. It can result in satisfying generalized Kerker condition and suppression of reflection from the array [38]. This opens up the possibility for more efficient control of reflective and transmissive properties of the arrays with a variety of multipoles [49,50], resonant suppression of reflection and transmission increase, as well as designs of perfect absorbers for light-harvesting devices.

4. In the realistic experimental structure and the arrays of a finite number of particles, the coupling between particle multipole moments can be stronger and result in a lattice resonance in a broad spectral range because of the boundary effects [37]. It has been shown recently that periodic arrays with about 50 particles arranged in square lattice enable not only well-pronounced lattice resonances but also provide an opportunity for their overlap with particle resonances resulting in resonant lattice Kerker effect [40].

In the model above, we have considered waves scattered by the particles in the uniform surrounding medium. One can speculate that the periodic array of nanoholes in the film may exhibit properties similar to the nanoparticle array. While each hole can be considered as scattering element and counterpart of nanoparticle scatterer, the system that includes nanoholes has a much higher level of complexity. For the thin films, it involves free-space waves propagating in the ambient medium on top and bottom sides of the film, surface waves propagating on each interface of the film and ambient medium, as well as other waves that may be supported by the film, such as guided or evanescent within the bulk part of the film [51-53]. This type of system requires a separate detailed study and may be a topic of the future investigation.

**Acknowledgments.**

This material is based upon work supported by the Air Force Office of Scientific Research under Grant No. FA9550-16-1-0088. The numerical studies have been supported by the Russian Science Foundation (Russian Federation), the project 16-12-10287.